\documentclass[prl,twocolumn,a4paper,superscriptaddress]{revtex4}
\usepackage{graphicx,amsmath,bm}
\usepackage{color}
\usepackage[normalem]{ulem}

\definecolor{nred}{rgb}{0.7,0.2,0.2}
\definecolor{nblack}{rgb}{0,0,0}
\definecolor{nblue}{rgb}{0.2,0.2,0.8}
\definecolor{ngreen}{rgb}{0.2,0.6,0.2}

\begin{document}

\title{Heralded Mapping of Photonic Entanglement into Single Atoms in Free Space:\\
Proposal for a Loophole-Free Bell Test}
\date{\today}
%\pacs{32.80.Qk}
\author{Nicolas Sangouard}
\author{Jean-Daniel Bancal}
\affiliation{Group of Applied Physics, University of Geneva, 1211 Geneva 4, Switzerland}
\author{Philipp M\"uller}
\affiliation{Experimentalphysik, Universit\"at des Saarlandes, Campus E\,2\,6, 66123 Saarbr\"ucken, Germany}
\author{Joyee Ghosh\footnote{Current Address: Department of Physics, Indian Institute of Technology, New Delhi, India}}
\affiliation{Experimentalphysik, Universit\"at des Saarlandes, Campus E\,2\,6, 66123 Saarbr\"ucken, Germany}
%\affiliation{Joint Quantum Institute, National Institute of Standards and Technology and University of Maryland, Gaithersburg, Maryland, USA}
\author{J\"urgen Esch\-ner}
\affiliation{Experimentalphysik, Universit\"at des Saarlandes, Campus E\,2\,6, 66123 Saarbr\"ucken, Germany}

\begin{abstract}
An obvious way to entangle two atoms located at remote locations is to produce a pair of
entangled photons half-way between the two atoms, to send one photon to each location and
to subsequently map the photonic entanglement into the atoms. The efficiency of this
process is, however, fundamentally limited due to overall transmission losses. We propose
a method to herald the success of the mapping operation in free space without destroying
nor revealing the stored quantum state. Interestingly for a Bell test, the heralding
signal does not open the detection loophole provided the measurement choice is performed
once the heralding is obtained only. We show through a detailed feasibility study that
this approach could provide an attractive alternative to Bell tests where the atom--atom
entanglement is created from atom--photon entanglement using an entanglement swapping
operation.
\end{abstract}
\maketitle

%%%%%%%%%%%%%%%%%%%%%%%%%%%%%%%%%%%%%%%%%%%%%%%%%%%%%%

\paragraph{Motivation}
What can be more fascinating than the violation of a Bell inequality? Yet, the Bell game
is simple, at least in principle. Two protagonists, Alice and Bob, share pairs of
entangled particles. Each of them randomly chooses measurements, $x$ and $y$
respectively, among an appropriate set of two projectors ($\{x=0,x=1\}$ and similarly for
$y$) and store the corresponding binary results, $a$ and $b$ ($\{a=+1,a=-1\}$ and
similarly for $b$). They repeat the experiment several times until they can estimate the
conditional probability distribution $p(ab|xy).$ The test really becomes exciting if the
measurement results violate a Bell inequality, e.\,g.\ \cite{Clauser69} if
\begin{equation}
\label{CHSH} \sum_{x,y=0}^1 (-1)^{xy} \Big(p\left(a=b|xy\right)-p\left(a\neq
b|xy\right)\Big) >2.
\end{equation}
In this case, Alice and Bob are forced to conclude that the observed correlations are
nonlocal, i.\,e.\ they got correlated results that are locally random, but cannot be
reproduced by a shared classical randomness. All the Bell experiments realized so far
point to the conclusion that this non-locality is, indeed, an element of the physical
reality, but they were all subjected to loopholes.

There are basically two loopholes, the detection loophole and the locality loophole. The
latter is closed if the measurement choice on Alice's side and the measurement result on
Bob's side, and vice versa, are spacelike separated. If this condition is not fulfilled,
the particles could simply communicate the measurement settings they experience to choose
the results accordingly. The former is related to the inefficiency of detections. The
particles could take advantage of the undetected events to answer when the measurement
settings are in agreement with a predetermined strategy only. The locality loophole was
addressed in experiments with entangled photons \cite{loc_loophole} and the detection
loophole with ions, atoms, and photons \cite{det_loophole, Hofmann12, vienna}. Closing
both in a single experiment would not only be the end of a long history of disputes, but
like many fundamental findings, it would open the way to fascinating applications. For
example, closing the detection loophole over tens of km would provide unique
opportunities to quantum-key-distribution protocols where the security does not rely on
the device that is used to generate the key \cite{diqkd}.

\paragraph{State of the art}
Photons are naturally suited for closing the locality loophole. They are fast, easy to
guide and can be produced at high repetition rates. However, the overall detection
efficiency has to be higher than 82.8\,\% if one wants to close the detection loophole
from the inequality (\ref{CHSH}), the so-called Clauser--Horne--Shimony--Holt (CHSH)
inequality \cite{Clauser69}, with maximally entangled states. Considering realistic
noise, achievable coupling into optical fibers and detection efficiencies, one rapidly
becomes aware that closing the detection loophole with photons between spacelike
separated locations is a very challenging task. Some looked for specific states or
peculiar Bell inequalities offering a better resistance to inefficiencies
\cite{loopholefree_photons}. Other studied the possibility of photonic Bell tests with
homodyne measurements to overcome the problem of the single-photon detection inefficiency
\cite{loopholefree_homodyne}. But the most promising approach for a loophole-free Bell
test uses atom--photon entanglement  \cite{atom_photon_th}, the photon allowing for the
distribution of entanglement over long distances and the state of an atom being detected
with an efficiency close to one. A lot of experimental effort has been devoted to the
creation of a single photon from a single atom where the photon polarization is entangled
with internal states of the atom \cite{Moehring04, Wilk07, Rosenfeld08}. Such
entanglement has further been used to entangle remote atoms through an entanglement
swapping operation \cite{Moehring07, Rempe, Hofmann12}. Hopefully, these impressive
experimental results could lead to the first loophole-free Bell test in a near future.

Experimental activities investigating the resonant interaction in free-space of a single
atom with single photons produced through the spontaneous parametric down conversion
(SPDC) have emerged in parallel \cite{Haase2009, Piro2009, Schuck2010}. In
ref.~\cite{Piro2011}, the interaction of single heralded SPDC photons with a single atom
has been demonstrated and in ref.~\cite{Huwer2012}, the possibility of obtaining
atom--photon entanglement from the absorption by a single atom of a photon belonging to a
polarization-entangled SPDC pair has been shown. One of the next great challenges could
naturally be the creation of entanglement between remote atoms, by producing a pair of
entangled photons half-way between two atoms and subsequently mapping the photonic
entanglement into the atoms. The efficiency of this process is, however, fundamentally
limited due to transmission losses. The mapping efficiency further decreases the
entanglement creation rate. Ref.~\cite{Lloyd01} proposed a way to herald the success of
the entanglement creation by exciting a cycling transition and by detecting the resulting
fluorescence detection. This heralding method has been implemented in
ref.~\cite{Kurz2012}. By further embedding the atoms in high-finesse cavities, the
authors of ref.~\cite{Lloyd01} end up with an efficient yet technologically demanding
architecture for quantum networking. We here focus on free-space interaction and propose
a fast and simple alternative method to herald the success of the mapping process without
revealing the stored quantum state. Although the proposed heralding process is
probabilistic, we show that it does not open the detection loophole provided that the
heralding signal is obtained before the measurement choice. We believe that the proposed
scenario is a potential candidate for the first loophole-free Bell test.

%%%%%%%%%%%%%%%%%%%%%%%%%%%%%%%%%%%%%%%%%%%%%%%%%%%%%%

\paragraph{Heralded mapping of photonic entanglement into single atoms: Principle}
\begin{figure}[ht!]
\includegraphics[width=0.9\linewidth]{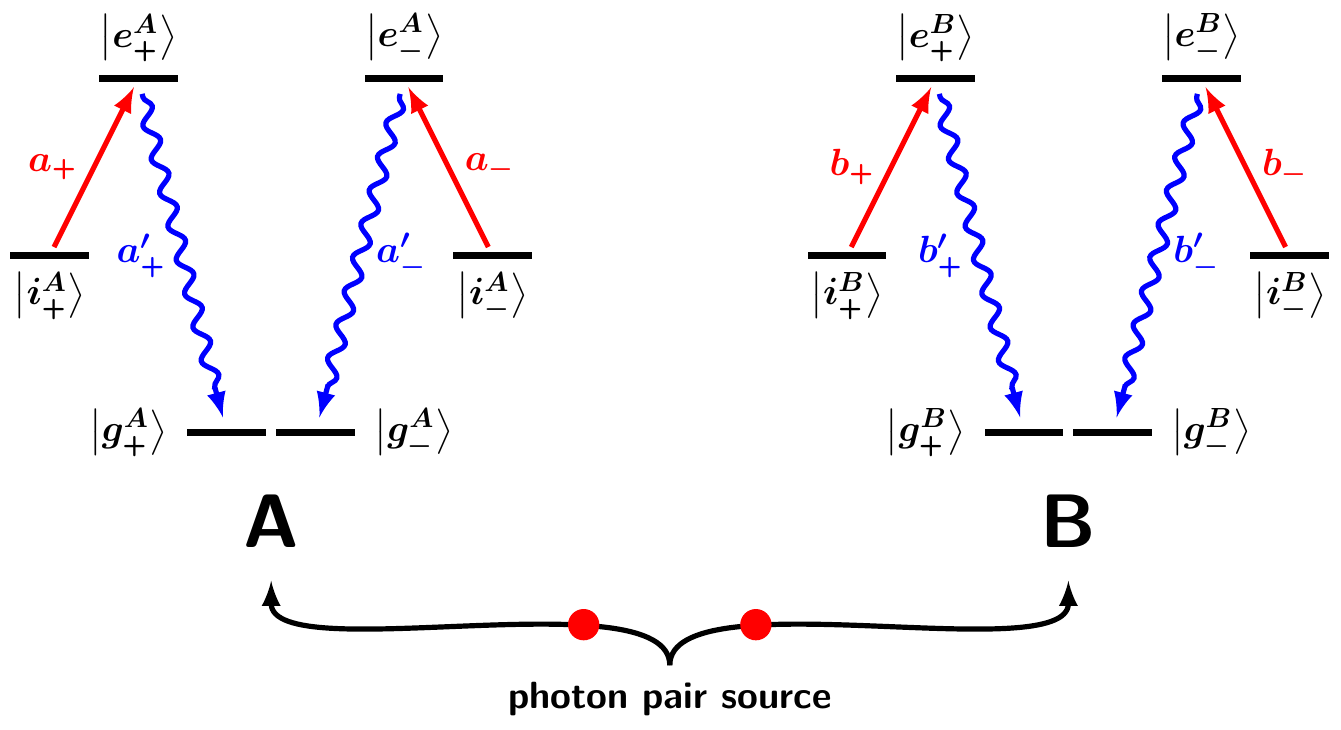}
\caption{Mapping polarization entanglement into the internal states of two atoms. The
success of the absorption process is heralded locally by the detection of a single photon
emitted spontaneously from an excited state. By choosing properly the detection basis,
the heralding signal does not reveal the polarization of the absorbed photon and leads to
the heralded creation of two entangled remote atoms. (See the main text for
details).\label{fig1}}
\end{figure}
Consider the scenario presented in fig.\,\ref{fig1} where two atoms, located at remote
locations A and B, contain each a double $\Lambda$-system of levels. Further consider
that they are initially prepared in a coherent superposition of two Zeeman levels
\begin{equation}
\label{ini_atom}
\psi_{i,+}^{\text{at}} = \frac{1}{2}\left(|i_+^A\rangle +
|i_-^A\rangle\right)\otimes \left(|i_+^B\rangle +
|i_-^B\rangle\right).
\end{equation}
A photon pair source at a central station located half-way between two atoms is excited
such that with a small probability $p,$ an entangled pair is created, corresponding to a
state
\begin{equation}
\Big[1+\sqrt{\frac{p}{2}}\underbrace{\left(a_+^{\dagger}b_-^{\dagger} -
a_-^{\dagger}b_+^{\dagger}\right)}_{\sqrt{2}\psi_-^{\text{ph}}}+\mathcal{O}(p) \Big] |0\rangle.
\end{equation}
Here, $a_+$ and $a_-$ ($b_+$ and $b_-$) are bosonic operators associated to two
orthogonal polarizations propagating towards Alice's (Bob's) location, e.\,g.\ in optical
fibers, and $|0\rangle$ is the vacuum state. The $\mathcal{O}(p)$ term introduces errors
in the protocol, leading to the requirement that $p$ has to be kept small enough, cf.\
below. If a pair is created, the corresponding two photons can both be absorbed when they
reach their destinations. Two successful absorptions transfer Alice's and Bob's atoms in
excited states and map the photonic entanglement into an atomic entanglement
$\psi_{\text{abs},-}^{\text{at}} = \frac{1}{\sqrt{2}}\left(e_+^Ae_-^B - e_-^Ae_+^B
\right).$ In principle, the maximum achievable probability $p_{\text{abs}}$ for a twofold
absorption is equal to $\frac{1}{4} p\eta_t^2,$ with $\eta_t$ being the transmission
efficiency from the source to one of the atoms. The pre-factor $\frac{1}{4}$ comes from
the fact that out of the initial state~(\ref{ini_atom}), on average only every second
photon can be absorbed. The coupling into the fiber $\eta_c$ and the absorption
efficiency $\eta_{\text{abs}}$ further limit $p_{\text{abs}}$ to $\frac{1}{4} p \eta_c^2
\eta_t^2 \eta_{\text{abs}}^2.$

Once they are excited, the atoms can spontaneously decay into ground states $g_+^A$ or
$g_-^A$ ($g_+^B$ or $g_-^B$) by emitting a photon with the corresponding polarization
$a_+'$ or $a_-'$ ($b_+'$ or $b_-'$), i.\,e.
\begin{equation}
\label{atom_photon}
\psi_{\text{em},-}^{\text{at}} = \frac{1}{\sqrt{2}} \left(g_+^{A}g_-^{B}a_+'^\dag b_-'^\dag -
g_-^{A}g_+^{B}a_-'^\dag b_+'^\dag \right)|0\rangle.
\end{equation}
Hence, the detection of one spontaneous photon at each location serves as a heralding
signal for the success of the mapping process and moreover, the entanglement is preserved
if the re-emitted photons are detected in the appropriate basis. For example, the
detection of two photons, one with the polarization
$a_H'=\frac{1}{\sqrt{2}}\left(a_+'+a_-'\right)$ and the other with
$b_H'=\frac{1}{\sqrt{2}}\left(b_+'+b_-'\right)$, projects the state of the atom pair into
\begin{equation}
\label{singlet_atomatom}
\psi_{\text{herald},-}^{\text{at}}=\frac{1}{\sqrt{2}}\left(g_+^{A}g_-^{B}-g_-^{A}g_+^{B}\right).
\end{equation}
The probability to obtain the heralding signal after absorption is $\eta_d^2$ where
$\eta_d$ is the efficiency with which a spontaneous photon is detected from a single
atom. Since twofold detection in $\{a_V', b_H' \},$ $\{a_H', b_V' \},$ $\{a_V', b_V' \}$
also projects the two atoms into $\psi_{\text{herald},-}^{\text{at}}$ (up to a unitary),
the overall efficiency for the twofold heralding is given by
$p_{\text{herald}}=\frac{1}{4} p \eta_c^2 \eta_t^2 \eta_{\text{abs}}^2 \eta_d^2.$

After the detection of the heralds, Alice and Bob can choose their measurement setting,
i.\,e.\ they perform a rotation on the two level system $\{g_+, \, g_-\}$ before
measuring the state of their atom through state-selective fluorescence (electron
shelving) or ionization. To close the locality loophole, it is important that the
distance $L$ separating Alice and Bob is such that $L/c$ (where $c$ is the velocity of
light in vacuum) is larger than the time it takes to know the atomic state once the
measurement setting is chosen. Independent of this is the other characteristic time
scale, i.\,e.\ the time it takes to receive a twofold herald, which can be very long.

We emphasize that the heralding process does not open the detection loophole if Alice and
Bob choose the measurement settings $x$ and $y$ only after the spontaneously emitted
photon is detected. This simply reduces to a pre-selection and none of the inefficiencies
mentioned so far enters in the detection efficiency required to close the detection
loophole. In particular, contrarily to the situation without pre-selection, there is no
limitation (other than technical ones) on the efficiency with which the photon states are
mapped to the atoms, if one is willing to lower the atomic entanglement-creation rate
\cite{Massar02}. Note also that Alice does not need to know whether Bob got the heralding
signal when she chooses her measurement setting. The detection of one spontaneous photon
at each location decides that a given run is going to contribute to the data of the CHSH
inequality test, and the measurement settings can be determined locally from the
polarizaton of the detected photon.

\begin{figure}[th!]
\includegraphics[width=0.9\linewidth]{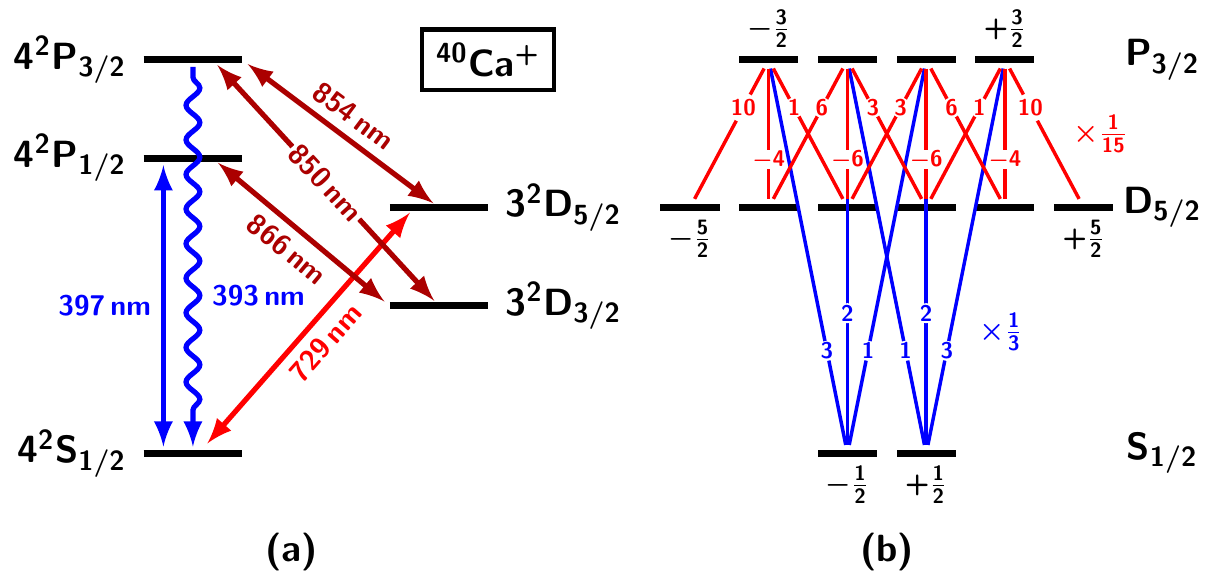}
\caption{(a) Level scheme and transitions for the $^{40}$Ca$^+$ ion. The branching ratios
for the decay of P$_{3/2}$ are 94\,\% into S$_{1/2}$, 6\,\% into D$_{5/2}$, and $<$\,1\,\% into
D$_{3/2}$. (b) Clebsch--Gordan coefficients (CGC); the CGC of a particular $m \rightarrow
m'$ transition is obtained by multiplying the modulus of the respective number with the
factor on the right, taking the square root, and applying the sign indicated with the
number.\label{fig2:CaLevels}}
\end{figure}

\paragraph{Heralded mapping of photonic entanglement into single atoms: Practical implementation}
We now discuss a practical implementation of the heralded entanglement distribution and
Bell test in more detail. For concreteness, we consider implementing the scheme using two
distant single trapped and laser-cooled $^{40}$Ca$^+$ ions. The $^{40}$Ca$^+$ ion is the
only single atomic system so far which has been coupled to entangled SPDC
photons~\cite{Schuck2010, Piro2011, Huwer2012}. A scheme of its relevant atomic levels
and transitions is shown in fig.\,\ref{fig2:CaLevels}. Based on the experimental work
reported in \cite{Haase2009, Piro2009, Schuck2010, Piro2011, Huwer2012} we assume that
the entangled photons are created at 854\,nm, resonant with the transition from the
metastable D$_{5/2}$ level to P$_{3/2}.$ Photon loss in optical fibers at this wavelength
is of the order of 1\,dB/km. This means that for $L=3$\,km, both photons will reach
Alice's and Bob's locations with a probability $\eta_t^2 \approx 0.5.$ This also
translates into an upper bound for the time delay between the measurement choice and the
measurement result of $10\,\mu$s. The heralding photons are assumed to be emitted on the
P$_{3/2}$ to S$_{1/2}$ transition at 393\,nm. The other levels and transitions presented
in fig.\,\ref{fig2:CaLevels} are employed for state preparation and
detection~\cite{Kurz2012}.

A weak magnetic field $\vec{B}$ defines a quantization axis and thereby, together with
the direction of propagation of the incoming photons $\vec{k}$, possible polarization
bases \cite{Huwer2012}: when $\vec{k} \parallel \vec{B}$, absorption of circularly
polarized photons leads to $\Delta m = \pm 1$ transitions; when $\vec{k} \perp \vec{B}$,
photons with polarization $\vec{\epsilon} \parallel \vec{B}$ induce $\Delta m = 0$
transitions, and photons with $\vec{\epsilon} \perp \vec{B}$ drive a superposition of the
two $|\Delta m| = 1$ transitions.

The initial state (\ref{ini_atom}) is implemented by preparing each atom in a coherent
superposition of sublevels $|\text{D},\pm\frac{5}{2}\rangle$ in D$_{5/2}.$ This is
achievable, for example, by coherent excitation on the S$_{1/2}$ to D$_{5/2}$ quadrupole
transition used as optical qubit in quantum logic experiments.
% refs. of the Blatt group
More precisely, this superposition can be produced starting from S$_{1/2}$ after cooling
the ion and optical pumping into a single $|\text{S},m\rangle$ sublevel, by employing a
Rabi $\pi/2$-pulse on $|\text{S},-\frac{1}{2}\rangle \to |\text{D},-\frac{5}{2}\rangle,$
followed by a Rabi $\pi$-pulse on $|\text{S},-\frac{1}{2}\rangle \to
|\text{S},+\frac{1}{2}\rangle$ (at radio frequency) and a subsequent Rabi $\pi$-pulse on
$|\text{S},+\frac{1}{2}\rangle \to |\text{D},+\frac{5}{2}\rangle.$ The whole state
preparation requires less than $25\,\mu$s.

Then, the ions are exposed to the photons at 854\,nm. The final state (\ref{atom_photon})
after emission of the heralding photon involves the magnetic sublevels of the S$_{1/2}$
state. Faithful mapping of the absorbed photon to the atomic state requires that the two
emission pathways be indistinguishable. Since they happen on transitions with different
$\Delta m$, the corresponding photons have to be projected on the same emitted
polarization $\vec{\varepsilon}_{393},$ as mentioned in the previous section. Detection
in this case is optimal along the quantization axis in a linear polarization basis; by
using two detectors for two orthogonal polarizations, the highest possible efficiency
$\eta_d$ will be obtained.

Current experiments \cite{Kurz2012, Schug2013} achieved the values $\eta_{\text{abs}}=2.5
\cdot 10^{-4}$ and $\eta_d=1.55\,\%$ (using a photon collection of 5.5\,\% and photon
detectors with an efficiency of 28\,\%). By using parabolic mirrors \cite{Maiwald2009} to
enhance the coupling to near unity, a maximal value of $\eta_\text{abs} \approx 6\,\%$
can be achieved (limited by the oscillator strength of the transition, cf.\
fig.\,\ref{fig2:CaLevels}) while $\eta_d$ can be of the order of 30\,\%. Furthermore we
assume that $\eta_c=70\,\%.$ All these numbers together yield an efficiency for the
twofold heralding of the order of $p_\text{herald} \approx 2 \cdot 10^{-5}p$.

\paragraph{Expected atom--atom entanglement visibility}
Taking the main experimental limitations into account, we now estimate the fidelity of
the heralded atom--atom entangled state with respect to the singlet state
(\ref{singlet_atomatom}). Our approach starts with the emission of photon pairs. Due to
possible multiple pair emission inherent in SPDC processes, the photon--photon entangled
state can be written $ \rho^\text{ph} =
V^{\text{ph}}|\psi_-^{\text{ph}}\rangle\langle\psi_-^{\text{ph}}|+(1-V^{\text{ph}})\frac{\mathbf{1}}{4}
$ where $\mathbf{1}$ stands for the identity and $V^{\text{ph}}=\frac{1-p/2}{1+p/2-p^2/2}
\approx 1-p+\mathcal{O}(p^2)$ \cite{Sekatski12} is the visibility of the interference that
would be obtained if Alice chooses the measurements $\sigma_x$ for example and Bob
rotates his measurement basis in the $\{xz\}$ plane. Additional errors (with
corresponding probability $e$) occurring at the further stage of the experiments degrade
the visibility according to $V \rightarrow (1-e)V.$ Once a photon pair is created, it
propagates to Alice's and Bob's locations. We assume that the polarization is actively
controlled such that the error on the polarization is of the order of
$e_{\text{pol}}=1\,\%$. The photonic state is then mapped to the atoms. The heralded
mapping operation (one photon to one atom) is estimated to be inaccurate at the same
order, $e_{\text{map}}=1\,\%.$ Dark counts of the heralding detectors add negligible
noise. The dominant error happens when one detector clicks because of a photon detection
but the other one produces a dark count. The probability of this erroneous event is given
by $p_{\text{dark}}=p \eta_c \eta_t \eta_{\text{abs}} \eta_d \eta_{\text{dark}}$ where
$\eta_{\text{dark}}=R_{\text{dark}} \Delta t,$ $R_{\text{dark}}$ being the dark count
rate (assumed to be 30 counts per second) and $\Delta t$ is the coincidence window (up to
20\,ns with respect to the photon coherence time of 7\,ns). The corresponding error is
given by $e_{\text{dark}}=\frac{p_{\text{dark}}}{p_{\text{dark}}+p_{\text{herald}}} \leq
10^{-3}.$ The resulting atom--atom entangled state is thus expected to be of the form
\begin{equation}
\rho^\text{at} =
V^{\text{at}}|\psi_{\text{herald},-}^{\text{at}}\rangle\langle\psi_{\text{herald},-}^{\text{at}}|
+ (1-V^{\text{at}})\frac{\mathbf{1}}{4}
\end{equation}
where the visibility of the atomic entanglement is given by $V^{\text{at}} =
V^\text{ph}(1-e_{\text{pol}})(1-e_{\text{map}})^2(1-e_{\text{dark}}) \approx
0.97(1-p)+\mathcal{O}(p^2).$

\paragraph{Performing the Bell test}
Once the heralding signal is obtained, Alice (or Bob) needs to be able to detect the
state of her (his) ion in various bases. The necessary rotations between
$|\text{S},+\frac{1}{2}\rangle$ and $|\text{S},-\frac{1}{2}\rangle$ are performed in up
to 10\,$\mu$s by a magnetic field at radio frequency. The detection then proceeds by
electron shelving of one $\text{S}_{1/2}$ sublevel into $\text{D}_{5/2}$ and measuring
resonance fluorescence from the $\text{S}_{1/2}  \leftrightarrow \text{P}_{1/2}$
transition. Following ref.~\cite{Myerson08}, such a measurement takes in average
145\,$\mu$s with a photon collection of 0.22\,\% and the mean accuracy of this procedure
was experimentally determined to be 99.99\,\%. However, assuming a global detection
efficiency of $\eta_d=30\,\%$ as before, the measurement time reduces to the time it
takes to perform to the local rotation (10\,$\mu$s). This is fast enough to close the
locality loophole with a distance of 3\,km. The measurement accuracy could realistically
be of $1-e_{\text{det}}=99.95\,\%.$

\paragraph{Expected violation of the CHSH inequality}
The resulting CHSH value is expected to be given by
\begin{equation}
S_{\text{exp}} = 2\sqrt{2}\,V
\end{equation}
with the statistical uncertainty
\begin{eqnarray}
\nonumber
&&\Delta S_{\text{exp}}=\frac{1}{\sqrt{2N}} \times
\\
\nonumber
&&\sqrt{3(1-\frac{1}{\sqrt{2}}V)^2(3+\frac{1}{\sqrt{2}}V)+(1+\frac{1}{\sqrt{2}}V)^2(3-\frac{1}{\sqrt{2}}V)}
\end{eqnarray}
where $V$ stands for $V^{\text{at}} (1-2 e_{\text{det}})^2$ \cite{Rosenfeld09}. In
principle, the value of $p$ needs to be optimized since a high CHSH value favors
$p\approx 0$ whereas a high heralding probability favors $p \approx 1.$ However,
practical considerations limit the pair production to approximately $5 \cdot 10^{5}$ per
seconds, i.\,e.\ $p=4 \cdot 10^{-3}$ in a coincidence window of $\Delta t = 7$\,ns. This
leads to $V^{\text{at}} \approx 96\,\%$ and $S_{\text{exp}}\approx 2.73$ so that $N=65$
events are necessary to conclude about the violation of the CHSH inequality with a
confidence level above 99.7\,\% (3 standard deviations). The duration of the state
preparation ($T \approx 25\,\mu$s) mainly defines the repetition rate and this translates
into an overall acquisition time of $N T / p_{\text{herald}} \approx 6$~hours and 30
minutes.

In order to rule out local models as plausible explanations of an observed Bell
inequality violation, it is convenient to estimate the probability that the observed
statistics be produced by such a model~\cite{Gill03,Zhang11}. Since the deviation of the
Bell values that can be obtained by local models in presence of finite statistics need
not be described appropriately by the experimental uncertainty $\Delta S_{\exp}$, we rely
on the following bound~\cite{Zhang11}:
\begin{equation}
P(S_{\text{exp}}|\text{local model}) \leq \exp\left(-N(S_{\text{exp}}-2)^2/32\right).
\end{equation}
This quantity is bounded by $0.05$ for the above value of $S_{\text{exp}}$, whenever
$N>181$. A clear demonstration that the observed statistics are not the result of a local
model is thus possible in a bit more than 18~hours.

\paragraph{Conclusion}
Our proposal opens a way to test Bell's inequalities in a loophole-free realization.
Specifically, it offers an interesting alternative to Bell tests where atom--atom
entanglement is created by means of an entanglement swapping operation. The latter is
being pursued by several groups and recent advances are very promising \cite{Hofmann12}.
Time will tell which one will allow one to answer a question lively debated about
non-locality. From a more applied perspective, our proposal may find applications in
quantum key distribution, either for implementing more secure protocols \cite{diqkd} or
for extending quantum key distribution over thousands of kilometers using quantum
repeaters \cite{Lloyd01, network_atom, QRep}. This supposes, however, significant
efficiency improvement in order to reach interesting key distribution rates.\\

\textit{Note that a related work has been carried out independently by Brunner et al., see arXiv:1303.6522.}

\paragraph{Acknowledgments}
We are grateful for discussions with Christoph Clausen and Pavel Sekatski. We acknowledge
support by the BMBF (QuOReP project, QScale Chist-ERA project) and the Swiss NCCR QSIT.

\end{document}